\documentclass[3p,times]{elsarticle}
\usepackage{lineno,hyperref}
\usepackage{graphicx}
\usepackage{dcolumn}
\usepackage{color}
\usepackage{bm}
\usepackage{epsfig}
\usepackage{graphicx}
\usepackage{psfrag}
\usepackage{amsmath,amssymb}
\usepackage{colordvi}
\usepackage{amsfonts}
\usepackage{enumerate}
\usepackage{slashed}
\usepackage{color}
\usepackage{xcolor}
\usepackage{soul}
\usepackage{ulem}
\newcommand*{\BibPath}{.}%

\newcommand{\eref}[1]{Eq.~(\ref{e.#1})}

\makeatletter
\def\ps@pprintTitle{%
\def\@oddfoot{}%
}
\makeatother

\begin{document}

\title{Unveiling the nucleon tensor charge  at Jefferson Lab: A study of the SoLID case}
\author[argonne,duke]{Zhihong Ye}
\ead{yez@anl.gov}
\author[jlab]{Nobuo~Sato}
\ead{nsato@jlab.org}
\author[mit]{Kalyan Allada}
\ead{kalyan@jlab.org}
\author[duke]{Tianbo Liu}
\ead{liutb@jlab.org}
\author[jlab]{Jian-Ping Chen}
\ead{jpchen@jlab.org}
\author[duke]{Haiyan Gao}
\ead{gao@phy.duke.edu}

\author[lanl]{Zhong-Bo Kang}
\ead{zkang@lanl.gov}
\author[psu,jlab]{Alexei~Prokudin}
\ead{prokudin@jlab.org}
\author[msu]{Peng~Sun}
\ead{psun@msu.edu}
\author[lbnl]{Feng~Yuan}
\ead{fyuan@lbl.gov}

\address[argonne]{Medium Energy Group, Physics Division, Argonne National Lab, Lemont, IL 60439, USA}
\address[duke]{Department of Physics, Duke University, Durham, NC 27708, USA}
\address[jlab]{Jefferson Lab, 
                   12000 Jefferson Avenue, 
                   Newport News, VA 23606, USA}
\address[mit]{Laboratory of Nuclear Science, Massachusetts Institute of Technology, Cambridge, MA 02139, USA}
\address[lanl]{Theoretical Division, Los Alamos National Laboratory, Los Alamos, New Mexico 87545, USA}
\address[psu]{Division of Science, 
                   Penn State Berks, 
                   Reading, PA 19610, USA}
\address[msu]{Department of Physics and Astronomy, Michigan State University, East Lansing, MI 48824, USA}
\address[lbnl]{Nuclear Science Division, 
                   Lawrence Berkeley National Laboratory, 
                   Berkeley, CA 94720, USA}

\begin{abstract}
Future experiments at the Jefferson Lab 12 GeV upgrade, in particular,
the Solenoidal Large Intensity Device (SoLID), aim at a very precise
data set in the region where the partonic structure of the nucleon is
dominated by the valence quarks. One of the main goals is to constrain
the quark transversity distributions. We apply recent theoretical
advances of the global QCD extraction of the transversity
distributions to study the impact of future experimental data from the
SoLID experiments. Especially, we develop a simple strategy based on the Hessian
matrix analysis that allows one to estimate the uncertainties of the
transversity quark distributions and their tensor charges extracted
from SoLID data simulation.  We find that the SoLID measurements with
the proton and the effective neutron targets can improve the precision
of the $u$- and $d$-quark transversity distributions up to one order
of magnitude in the range $0.05<x<0.6$.
\end{abstract}

\begin{keyword}
Quantum Chromo Dynamics;
Semi-Inclusive Deep Inelastic Scattering; 
Tensor charge; Transversity; 
Jefferson Lab 12 GeV Upgrade; 
SoLID;  
JLAB-THY-16-2328 
\PACS  12.38.-t, 13.85.Hd, 13.88.+e, 14.65.Bt
\end{keyword}

\maketitle

\section{Introduction}
%
The nucleon tensor charge is a  fundamental property  of the nucleon and
its determination is among the main goals of existing and future
experimental facilities~\cite{Ralston:1979ys,Jaffe:1991kp,Cortes:1991ja,
Barone:2001sp,Boer:2011fh,Dudek:2012vr,Accardi:2012qut}.  It also
plays an important role in constraining  new physics beyond the
standard model \cite{DelNobile:2013sia,Bhattacharya:2011qm,Courtoy:2015haa} and has been an active subject of lattice QCD
~\cite{Bhattacharya:2011qm,Green:2012ej,Bhattacharya:2013ehc,Bhattacharya:2016zcn,Chen:2016utp,Bali:2014nma,Gupta:2015tpa,Yamanaka:2015lfk,Aoki:2010xg,Gockeler:2005cj}
and  Dyson-Schwinger Equation (DSE)~\cite{Yamanaka:2013zoa,Pitschmann:2014jxa} calculations. In terms of the
partonic structure of the nucleon, the tensor charge, $\delta q$ for a
particular quark type $q$,  is constructed from the quark transversity
distribution, $h_1(x,Q^2)$, which is one of the three
leading-twist quark distributions that describe completely spin-1/2
nucleon~\cite{Ralston:1979ys,Jaffe:1991kp,Cortes:1991ja,Barone:2001sp,Boer:2011fh}:
\begin{eqnarray}
\delta q \left(Q^2\right) \equiv   \int_{0}^{1}dx \, \left(h_1^q(x,Q^2) - h_1^{\bar q}(x,Q^2)\right)
 \ .
\end{eqnarray}

It is extremely important to extend the experimental study of the
quark transversity distribution to both large and small Bjorken
$x$ to constrain the total tensor charge contributions.  The Jefferson
Lab 12 GeV program~\cite{Dudek:2012vr}  is going to explore the region
of relatively large-$x$ dominated by valence quarks while the
planned Electron Ion Collider
~\cite{Boer:2011fh,Accardi:2012qut,Aschenauer:2014twa} is going to
extend the range  to
unexplored lower values of $x$, providing a possibility to
study the anti-quark transversity distributions.

In this paper we analyze the impact of future proposed SoLID
experiment at Jefferson Lab 12 GeV on the determination of tensor charge and
transversity distributions for $u$- and $d$-quarks.
Our studies are based on the QCD global fit of the available
Semi-Inclusive Deep Inelastic Scattering (SIDIS) data and $e^+e^-$
annihilation into hadron pairs performed in Ref.~\cite{Kang:2015msa}
which we will refer as KPSY15. The current available experimental data suggests that anti-quark
transversities are very small compared to $u$- and $d$-quark transversities. In
this study we assumed that anti-quark transversities are
negligible. 
Using the best fit of transversity distributions of
Ref.~\cite{Kang:2015msa} we simulated pseudodata for SoLID experiment
and estimate  the improvement of $u$- and $d$-quark transversity
distributions with respect to our present knowledge. In order to perform a reliable estimate of improvement we develop a simple method based on Hessian error analysis described in Section~\ref{Impact}.

This study also provides information on contribution of tensor
charge from kinematical region of Jefferson Lab 12 GeV  and will serve as a
guide in planning future experiments.

\section{Present status of extraction of transversity from
experimental data\label{Present}}
Transversity 
is a chiral odd quantity and thus in order
to be measured in a physics process it should couple to another
chiral odd distribution.  There are  several ways
of accessing transversity. It can be studied  
in SIDIS process where
it couples, for instance, to the Collins TMD fragmentation
functions~\cite{Collins:1992kk}, and produces the so-called Collins
asymmetries. Transversity  
can  also couple to the dihadron interference fragmentation
functions in SIDIS \cite{Collins:1993kq} and thus collinear transversity
can be studied directly. Transversity can be studied in the Drell-Yan process in polarized
hadron-hadron scattering \cite{Barone:2005pu,Aschenauer:2015eha} where it couples either to anti-quark transversity or to the so-called the Boer-Mulders functions.

SIDIS experimental measurements have been made at  HERMES
~\cite{Airapetian:2004tw,Airapetian:2010ds},
 COMPASS
~\cite{Alekseev:2008aa, Adolph:2012sn,Adolph:2014zba}, and JLab HALL A~\cite{Qian:2011py}
experiments. The BELLE, BABAR and
the BESIII
collaborations have studied the asymmetries in $e^+e^-$
annihilation into hadron
pairs at  the center of mass energy around
$\sqrt{s}\simeq 10.6$
GeV~\cite{Abe:2005zx,Seidl:2008xc,Garzia:2012za}, and  $\sqrt{s}\simeq
3.6$ GeV~\cite{Ablikim:2015sma}, respectively.

The effort to extract transversity distributions and Collins
fragmentation functions has been carried out extensively in the last
few years~\cite{Anselmino:2007fs,Anselmino:2008jk,Anselmino:2013vqa,Anselmino:2015sxa,Kang:2015msa}.
QCD analysis of the data where transversity couples to the so-called dihadron
interference fragmentation functions was performed 
in Ref.~\cite{Radici:2015mwa}.
These results have demonstrated the powerful capability of the
asymmetry measurements in constraining quark transversity
distributions and hence the nucleon tensor charge in high energy
scattering experiments.  
The first extraction of the transversity  distributions and Collins
fragmentation functions with TMD evolution was performed in
Refs.~\cite{Kang:2014zza,Kang:2015msa}.  

 Collins
asymmetries in SIDIS are generated by the convolution of the
transversity function  $h_1$ and Collins function $H_1^\perp$.
The relevant contributions to the SIDIS cross-sections are
\begin{align}
\frac{d^5\sigma(S_\perp)}{dx_B dy dz d^2P_{T}}
= \sigma_0
\Big[F_{UU} +  
  \sin(\phi_h+\phi_s)\,
\frac{2 (1-y)}{1+(1-y)^2} \, F_{UT}^{\sin\left(\phi_h +\phi_s\right)} + ...\Big]\,,
  \label{eq:aut-collins}
\end{align}
where $\sigma_0 = \frac{2\pi \alpha_{\rm em}^2}{Q^2}\frac{1+(1-y)^2}{y}$, 
and  $\phi_s$ and $\phi_h$ are the azimuthal angles for the nucleon spin and the transverse momentum 
of the outgoing hadron with respect to the lepton plane, respectively. $F_{UU}$ and $F_{UT}^{\sin(\phi_h+\phi_s)}$ are the unpolarized and transverse spin-dependent 
polarized structure functions respectively, and the ellipsis represents other polarized structure functions not relevant for this analysis. 
The polarized structure function $F_{UT}^{\sin\left(\phi_h
+\phi_s\right)}$ contains the convolution of transversity  distributions with the
Collins \ fragmentation functions, $h_1 \otimes H_1^\perp$, and unpolarized structure function $F_{UU}$ is the convolution of  the unpolarized TMD distributions and
 the unpolarized  fragmentation functions, $f_1 \otimes D_1$.
The Collins asymmetry is defined as
\begin{align}
A_{UT}^{\sin\left(\phi_h +\phi_s\right)}(x, y, z, P_{T})
=  
\frac{2 (1-y)}{1+(1-y)^2} \, \frac{F_{UT}^{\sin\left(\phi_h +\phi_s\right)}}{F_{UU}}  \ .
  \label{eq:aut-collins1}
\end{align}

Neglecting sea quark contributions, the structure function 
$F_{UT}^{\sin\left(\phi_h +\phi_s\right)}$ for the proton (P)
and the neutron  (N) targets can be written as: 
\begin{align}
F_{UT}^{\sin\left(\phi_h +\phi_s\right)}(P,\pi^+) &= e_u^2 h_1^{u} \otimes H_1^{\perp, fav} +
e_d^2 h_1^{d} \otimes H_1^{\perp, unf}\; , \\
F_{UT}^{\sin\left(\phi_h +\phi_s\right)}(P,\pi^-) &= e_u^2 h_1^{u} \otimes H_1^{\perp, unf} +
e_d^2 h_1^{d} \otimes H_1^{\perp, fav}\; ,\\
F_{UT}^{\sin\left(\phi_h +\phi_s\right)}(N,\pi^+) &= e_u^2 h_1^{d} \otimes H_1^{\perp, fav} +
e_d^2 h_1^{u} \otimes H_1^{\perp, unf}\; ,\\
F_{UT}^{\sin\left(\phi_h +\phi_s\right)}(N,\pi^-) &= e_u^2 h_1^{d} \otimes H_1^{\perp, unf} +
e_d^2 h_1^{u} \otimes H_1^{\perp, fav}\;.
\end{align}
Here $H_1^{\perp, fav}$ and  $H_1^{\perp, unf}$, are  the {\it
favored} and the {\it unfavored} Collins fragmentation functions,
respectively. In this context, {\it favored} refers to fragmentation
of struck quarks of the same type as the constituent valence quarks of
the produced pion while the {\it unfavored} being the opposite case.
Previous global analysis \cite{Kang:2015msa,Anselmino:2013vqa} have
found that both the {\it favored} and {\it unfavored} Collins
functions have approximately similar magnitude (with opposite signs).
Therefore, since $e_u^2$ = 4$e_d^2$, 
the $u$-quark
transversity is more constrained in the proton sample than $d$-quark
transversity and the situation is reversed in the neutron case.  One
expects from these considerations that only the neutron target can
help to reach the same relative impact on determination of $d$-quark
transversity compared to improvement of $u$-quark transversity from
the proton target data.

 In the KPSY15 analysis the transversity distributions was parametrized as
at the input scale $Q_0=\sqrt{2.4}$ GeV as 

\begin{align}
h_1^{q}(x,Q_0)= & N_{q}^h x^{a_{q}}(1-x)^{b_{q}} \frac{(a_{q} + b_{q})^{a_{q} + b_{q}}}
{a_{q}^{a_{q}} b_{q}^{b_{q}}} \, \cdot
 \frac{1}{2}\left (f_{1}^q(x,Q_0) + g_{1}^q(x,Q_0)  \right ) \ ,
\end{align}
where $f_{1}^q$ and $g_{1}^q$ are the collinear
unpolarized~\cite{Lai:2010vv} and polarized~\cite{deFlorian:2009vb}
quark distributions for $q=u-$ and $d$-quark, respectively.

On the other hand, the twist-3 Collins fragmentation functions
 were parametrized in terms of the
unpolarized fragmentation functions,
\begin{align}
\hat{H}_{fav}^{(3)}(z,Q_0)&= N_{u}^c z^{\alpha_{u}}(1-z)^{\beta_{u}} D_{\pi^+/u}(z,Q_0) \ , \\
\hat{H}_{unf}^{(3)}(z,Q_0)&= N_{d}^c z^{\alpha_{d}}(1-z)^{\beta_{d}} D_{\pi^+/d}(z,Q_0) \ , 
\end{align}
which correspond to the favored and unfavored Collins fragmentation
functions, respectively.  
For $D_{\pi^+/q}$ we use the recent extraction from
Ref.~\cite{deFlorian:2014xna}.
 
 In summary, the analysis
of KPSY15  used a total of 13 parameters
in  their global fit: $N_u^h$, $N_d^h$, $a_u$, $a_d$, $b_u$, $b_d$,
$N_u^c$, $N_d^c$, $\alpha_u$, $\alpha_d$, $\beta_d$, $\beta_u$, $g_c$
(GeV$^2$), where $g_c$ is a parameter to model the width of the Collins fragmentation function. 
The parameters are shown in Table~\ref{parameters}.
\begin{table}[htb]
\centering
\begin{tabular}{l c l l c l l c l}
\hline
 & & & & & & & &\\
$N_u^h$ &=& $0.85\pm 0.09$ & $a_u$ &=& $ 0.69 \pm 0.04$ & $b_u$ &=& $ 0.05 \pm 0.04$ \\
$N_d^h$ &=& $-1.0\pm 0.13$ & $a_d$ &=& $ 1.79 \pm 0.32$ & $b_d$ &=& $ 7.00 \pm 2.65$  \\
$N_u^c$ &=& $-0.262\pm 0.025$ & $\alpha_u$ &=& $ 1.69 \pm 0.01$ & $\beta_u$ &=& $ 0.00 \pm 0.54$ \\
$N_d^c$ &=& $0.195\pm 0.007$ & $\alpha_d$ &=& $ 0.32 \pm 0.04$ & $\beta_d$ &=& $ 0.00 \pm 0.79$ \\
$g_c$ &=& $0.0236\pm 0.0007$&\multicolumn{3}{l}{(GeV$^2$)}\\
& & & & & & & &\\
\hline
\end{tabular}
\caption{Fitted parameters of the transversity   distributions for $u$- and $d$-quark, and Collins fragmentation functions. The table is from Ref.~\cite{Kang:2015msa}}
\label{parameters}
\end{table}

Since the existing experimental data have only probed the limited region $0.0065
< x < 0.35$, the following partial contribution to the tensor charge,
neglecting anti-quark contributions, was defined~\cite{Kang:2015msa}
\begin{eqnarray}
\delta q^{[x_{\rm min},x_{\rm max}]}\left(Q^2\right) \equiv   \int_{x_{\rm min}}^{x_{\rm max}}dx \, h_1^q(x,Q^2) \ .
\label{eq:trunctensor}
\end{eqnarray}
  
\section{Simulated Data for SoLID\label{Simulated}}
%
Several SIDIS experiments have been approved at Jefferson Lab 12 GeV  to
measure the asymmetries from proton and neutron targets with
polarization in both the transverse and longitudinal directions. 
Among those, three  Hall A experiments, E12-10-006~\cite{e1210006} (90
days), E12-11-007~\cite{e1211007} (35 days), and
E12-11-108~\cite{e1211108} (120 days), plan to take data using 
the proposed high
intensity and large acceptance device named SoLID~\cite{solid_pcdr,Chen:2014psa},
and measure both the single-spin asymmetries (SSA)
and double-spin asymmetries (DSA) on polarized $\mathrm{NH_{3}}$
(proton) and $\mathrm{^{3}He}$ (effective neutron) targets. These
experiments can produce an extensive set of SIDIS data with very high
accuracy and 
thus provide unique opportunity to study TMD structure
functions in the valence quark region.

In these experiments, the electron beam energy will be set at two
different values, 8.8\,GeV and 11\,GeV. The momentum of the detected
electrons and hadrons can range from 1~GeV/c up to their maximum
values. The SoLID configuration dedicated to the SIDIS measurements
provides a full $2\pi$ coverage in azimuthal angle and a coverage of
the polar angle from $8^{\circ}$ up to $24^{\circ}$.  The polarized 
luminosities of the proton target and the $\mathrm{^{3}He}$ target are
$10^{35}$ cm$^{-2} \cdot$ s$^{-1}$ and $10^{36}$ cm$^{-2}\cdot $
s$^{-1}$, respectively.  The polarization and dilution factor of the
proton ($\mathrm{^{3}He}$) target are 70\% (60\%) and 0.13 (0.3),
respectively. 

For the purpose of the present analysis, we simulate the Collins
asymmetries  using the KPSY15 parametrization at   the kinematic
settings presented in the proposals of these experiments ~\cite{e1210006,e1211007, e1211108}. The high luminosity 
allows us to bin the data in four dimensions, e.g. $x$, $z$, $Q^2$, and
$P_{T}$.  The acceptance of the proposed SoLID
measurements are summarized in Table~\ref{tab:kinem}.
There are in
total 1014 bins for $\mathrm{^{3}He}(e,e')\pi^{+}$, 879 bins for
$\mathrm{^{3}He}(e,e')\pi^{-}$, 612 bins for $p(e,e')\pi^{+}$, and 488
bins for $p(e,e')\pi^{-}$, respectively.
The number of events in each
bin is calculated by integrating over the cross sections and
acceptance of individual events in this bin, and then accounting for
the detector efficiencies and the target related characteristics, such
as the luminosity, target polarization, effective neutron polarization
as well as the dilution factor.  The average values of  $x$, $z$, $Q^2$, and $P_{T}$ are 
recorded  in each bin together with the statistical uncertainty. 

We also estimate the overall systematic uncertainty related to the experimental measurement,
such as the raw asymmetry, target polarization, detector resolution,
nuclear effects, random coincidence, and radiative corrections.
The final uncertainties of the simulated Collins asymmetries are
given as statistical and systematic uncertainties added in
quadrature.
 

\begin{table}[bth!]
\centering
\begin{tabular}{|c|c|c|c|c|}
\hline
Variable      & Min  & Max & Bin Size & Bins  \\ \hline
  $Q^{2}$    & 1.0~GeV$^{2}$  & 8.0~GeV$^{2}$  & $\sim 1.0$~GeV$^{2}$   & 6 bins\\
  $z$           &  0.3  & 0.7   & 0.05  & 8 bins\\
  $P_{T}$    &  0.0~GeV &  1.6~GeV&  0.2~GeV & $\leq$ 8 bins \\
  $x$           & 0.05 & 0.6 & NA  &  $\leq$ 8 bins \\\hline
\end{tabular}
\caption{\label{tab:kinem} Kinematic limits of SoLID. The bin-size for $P_T$ is doubled when number of total events $<5\times 10^{6}$, and the bin size in $x$ varies to keep number of events in one bin $\sim10^{6}$. The actual bin size of the last bin with the center at $x=0.6$ will extend up to $x \sim 0.7$.}
\end{table}
%
\begin{figure}[htbp]
\centering
\includegraphics[width=0.45\textwidth]{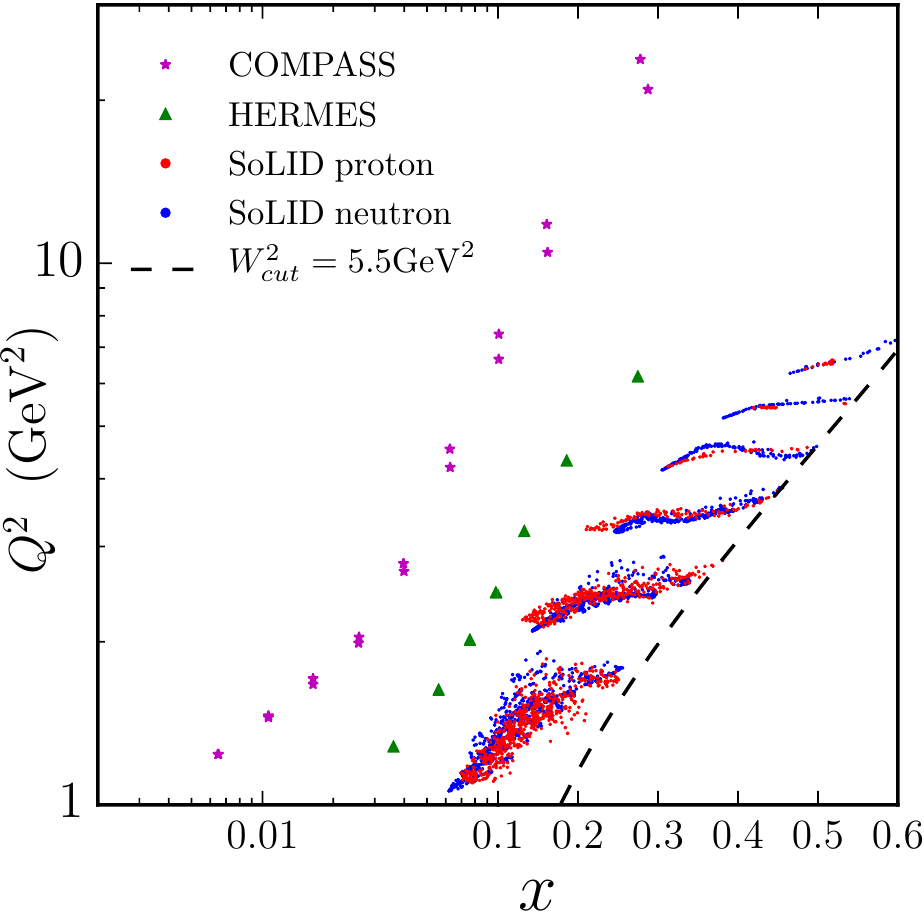}  
\vskip -0.3cm 
\caption{ $x-Q^2$ kinematical plane of bins for SoLID data with HERMES~\cite{Airapetian:2004tw,Airapetian:2010ds} and COMPASS~\cite{Alekseev:2008aa, Adolph:2012sn,Adolph:2014zba} data sets.}
\label{fig:kin}
\end{figure}

The distribution of bins in $x-Q^2$ plane for SoLID and the comparison to
HERMES~\cite{Airapetian:2004tw,Airapetian:2010ds} and
COMPASS~\cite{Alekseev:2008aa, Adolph:2012sn,Adolph:2014zba}, bins
are 
presented in Fig.~\ref{fig:kin}. The SoLID experiment plans to extend
mainly into the larger $x$ region with $Q^2$ coverage comparable with
HERMES. A direct comparison of the statistical precision of SoLID and
the existing data is not possible due to different binning criteria
between experiments, but an estimate of the level of precision can be
given. For example, the average statistical precision of each bin for
SoLID is about 1\% consisting of more than 600 bins for
$p(e,e')\pi^{+}$ channel, compared to 37.1\% (relative to the size of
the asymmetry)  for HERMES consisting of 7 bins in $x$ shown in
Fig.~\ref{fig:kin} for the same channel. Note that SoLID implements
$W^2$ cut at around 5.5 GeV$^2$. We leave the feasibility of implementing
target mass corrections and usage of low $W$ region in the analysis of
the experimental data for future developments of the theory and
phenomenology.

\section{Error estimation methodology from simulated data\label{Impact}}
 
    In this section we describe  the new method to estimate the impact of
the future SoLID data to the transversity distribution of $u$- and
$d$-quarks.  Our method follows Bayesian statistics where the new
information is added sequentially on top of the prior knowledge
without requiring   a combined analysis of the old data and
the new data.  We provide a simple strategy to quantify the impact of
new measurements on the transversity distribution using the Hessian
approach.

In general the information of the best fit parameters
and their uncertainties is encoded in the likelihood function
\begin{align}
\mathcal{L}(D|\bm{\alpha})
\sim\exp\left(-\frac{1}{2}\chi^2(\bm{a},D)\right)
\label{eq:likelihood}
\end{align}
where $\bm{a}$ represents a vector of  the model parameters and $D$ denotes
collectively the experimental data points and their uncertainties . $\chi^2$ is
the standard Chi-squared function defined as
\begin{align}
\chi^2=\sum_i \left(\frac{D_i-T_i(\bm{a})}{\delta D_i}\right)^2 \ ,
\label{eq:chi2}
\end{align}
where $T_i(\bm{a})$ is the theoretical calculation for experimental
measurement of $D_i$ and  $\delta D_i$ is the experimental error of the
measurement. The probability density of the parameters can be 
constructed from the likelihood function using the Bayes' theorem:
\begin{align}
\mathcal{P}(\bm{a}|D)\sim \mathcal{L}(D|\bm{a})~\pi(\bm{a}) \ ,
\end{align}
where $\pi(\bm{a})$ is the prior distribution.  Typically the latter is set  to be normalized theta functions to remove unphysical regions
in the parameter space. The expectation value and variance for an
observable $\mathcal{O}$ (i.e. $h_1^{u,d},\delta u,\delta d$) can be
estimated as 
\begin{align}
E[\mathcal{O}]&=\int d^n a\ \mathcal{P}(\bm{a}|D)\ \mathcal{O}(\bm{a}) \ ,
\notag\\
V[\mathcal{O}]&=\int d^n a\ \mathcal{P}(\bm{a}|D)\ [\mathcal{O}(\bm{a})-E[\mathcal{O}]]^2 \ .
\label{e.EV}
\end{align}
In most of the situations the evaluation of the above integrals are not
practical due to the large number of parameters needed in the model 
as well as numerical cost in evaluating  $\mathcal{P}(\bm{a}|D)$
or equivalently the $\chi^2$ function. A traditional method to
estimate \eref{EV} is the maximum likelihood (ML). 
First the parameters $\bm{a}_0$ that maximizes the likelihood (or minimized the $\chi^2$
function) is determined so that one can write 
\begin{align}
E[\mathcal{O}]\approx \mathcal{O}(\bm{a}_0) \ .
\label{e.EML}
\end{align}
 
A very simple method to estimate the variance is the Hessian approach
\cite{Stump:2001gu,Pumplin:2001ct}. 
The idea is to compute the
covariance matrix of the parameters using the Hessian of the $\chi^2$
function: 
\begin{align}
C_{i,j}^{-1}\approx H_{i,j}=\left. \frac{1}{2}\frac{\partial\chi^2(\bm{a},D)}{\partial
a_i\partial a_j}
\right|_{\bm{a}_0} \ .
\label{eq:H}
\end{align}
From the eigen values $\lambda_k$ and their corresponding normalized
eigen vectors $\bm{v}_k$ of the covariance matrix one can estimate 
the variance on $\mathcal{O}$ as 
\begin{align} 
V[\mathcal{O}]=\frac{\Delta \chi^2 }{4}\sum_k 
\left(
\mathcal{O}(\bm{a}_0+\sqrt{\lambda_k}\bm{v}_k)
-
\mathcal{O}(\bm{a}_0-\sqrt{\lambda_k}\bm{v}_k)
\right)^2.
\label{eq:estimate}
\end{align}
The factor of $\Delta \chi^2$ (commonly known as the tolerance
factor) is introduced in order to accommodate possible tensions among
the data sets. In the ideal Gaussian statistics, $68\%$ CL corresponds
to $\Delta \chi^2=1$. In the present analysis we use the value of $\Delta
\chi^2=29.7$ quoted in the KPSY15 analysis.  We stress however
that our analysis focuses on the relative improvement after inclusion
of the future SoLID data for which the tolerance factor drops out. 

A simple  Bayesian  strategy to estimate the impact of the future measurements on
the existing uncertainties is to update the covariance matrix. Since
the only information provided is the projected statistical and
systematic uncertainties, the expectation values (or equivalently
$\bm{a}_0$) remain the same. To update the covariance matrix we note
that the $\chi^2$ function is additive and one can write the new 
Hessian matrix as

\begin{align}
H^{\rm New}_{i,j}=
\left. \frac{1}{2}\frac{\partial\chi^2(\bm{a},D_{\rm old})}{\partial
a_i\partial a_j}
\right|_{\bm{a}_0}
+
\left. \frac{1}{2}\frac{\partial\chi^2(\bm{a},D_{\rm new})}{\partial
a_i\partial a_j}
\right|_{\bm{a}_0}  \ ,
\label{eq:Hnew}
\end{align}
where $D_{old}$ is the data set used in a previous analysis (i.e.
KPSY15) and the $D_{new}$ is the simulated data set for the future
experiment. In our analysis only the covariance matrix from the KPSY15
analysis was provided. The new covariance matrix with the projected
SoLID measurements was calculated as 
\begin{align}
C_{\rm New}^{-1}=
H^{\rm New}_{i,j}=
C_{\rm {\rm KPSY15}}^{-1}
+
\left. \frac{1}{2}\frac{\partial\chi^2(\bm{a},D_{\rm SoLID})}{\partial
a_i\partial a_j}
\right|_{\bm{a}_0} \ .
\label{eq:Hnew}
\end{align}
Using the new covariance matrix one can determine the impact of
future data sets by estimating the uncertainties
for the observables $\mathcal{O}$, such as transversity or tensor
charges, using Eq.~(\ref{eq:estimate}).

\section{Tensor charge and transversity from SoLID\label{Tensor}}
\begin{figure}
\centering
\includegraphics[width=0.95\textwidth]{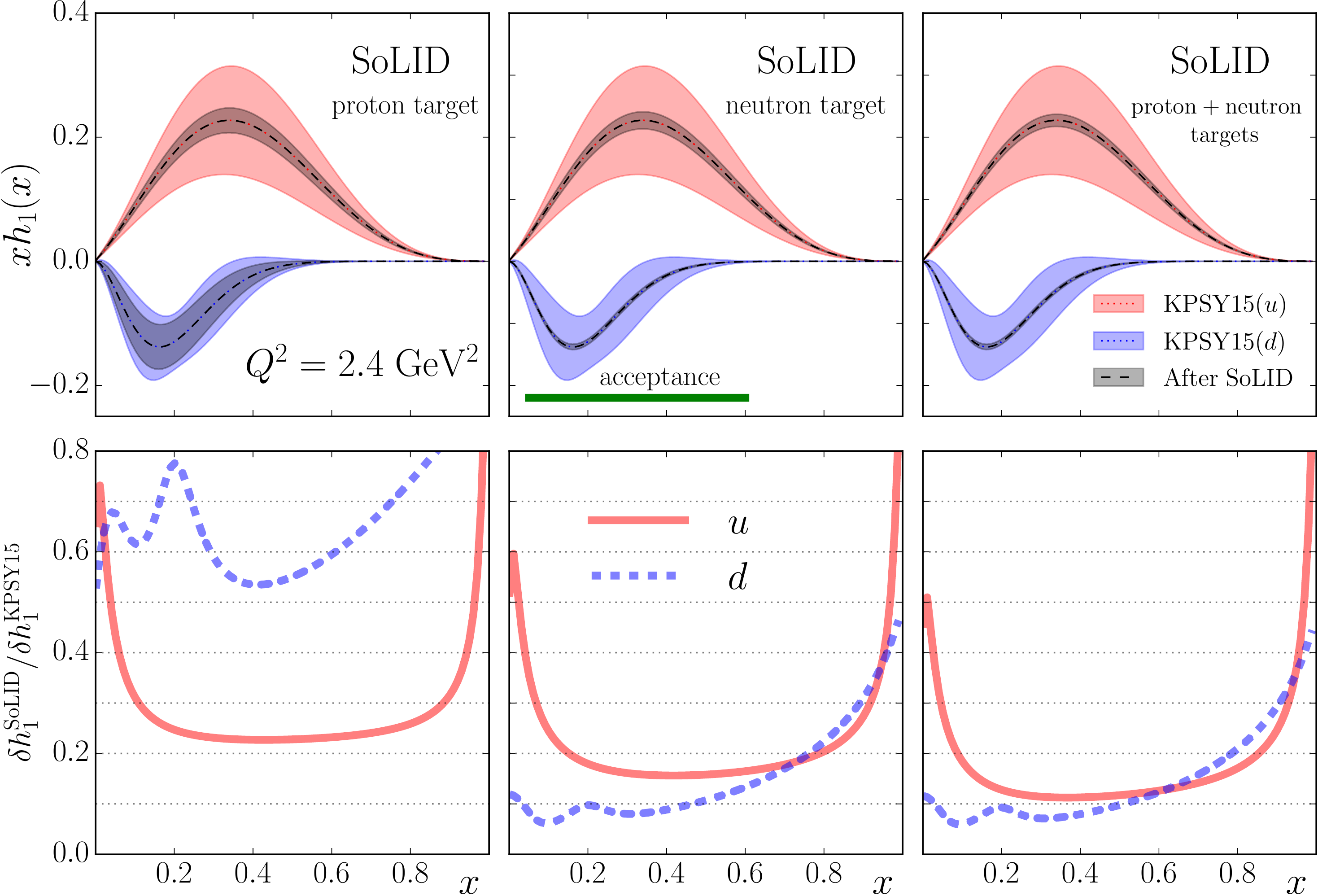}
\vskip -0.3cm 
\caption
{ 
{\bf Upper panels:} $u$-quark  and $d$-quark transversity
distributions at $Q^2=2.4$ (GeV$^2$) as a function of $x$ with
existing errors from KPSY15 (light shade area) and the
estimated errors after the SoLID data (both statistical and
systematical errors are included in quadrature) are taken into
account. The acceptance region in $x$ of the SoLID experiment is
indicated by the green horizontal line. Left plot: only the proton
target data are taken into account, central plot: only the neutron
target data are taken into account, right plot: combination of proton
and neutron targets data are taken into account. {\bf Bottom panels:}
The ratio of the estimated errors and the current errors of
transversity, $\delta h_1^{\rm SoLID}/\delta h_1^{\rm KPSY15}$, for $u$ (solid
line) and $d$ (dashed line) quarks.  Left plot: the proton target,
central plot: the neutron target, right plot: combination of proton
and neutron targets. The ``bumps" around $x\simeq 0.2$ of the $d$-quark
ratio plots are artifacts of usage of Soffer positivity bound~\cite{Soffer:1994ww} when parametrizing transversity.
}
\label{fig:transversity}
\end{figure}

Our results for $u$- and $d$-quark transversity distributions at
$Q^2 = 2.4$ GeV$^2$ are  presented in Fig.~\ref{fig:transversity} along
with results from KPSY15.  The uncertainties of KPSY15  are
 given 
as light shaded bands, while the projected errors after the SoLID data
are taken into account are  shown as dark shaded bands.  To quantify the improvement of 
adding the future SoLID data, we show in the bottom plots of
Fig.~\ref{fig:transversity} 
 the ratio of the estimated errors relative to the current errors. 
The results are shown 
using only the proton target data (left
panels), the neutron data (central panels), and combination of the
proton and the neutron data sets (right panels). In KPSY15 the
uncertainty bands for transversity was calculated using the {\it
envelop} method with a tolerance of $\Delta \chi^2 = 29.7$ which
differs somehow from our Hessian error analysis. We stress that while
the absolute error bands can differ depending on the error analysis,
the ratio of the errors is  independent of the error analysis.

One can see that, the proton target data improves $u$-quark transversity uncertainty  
 (as can be seen from the left plot of the
bottom panel of Fig.~\ref{fig:transversity}) while $d$-quark
transversity improvement remains at a modest $\sim 60$\% level.
The effective neutron target data as expected allows for a much better
improvement of $d$-quark transversity uncertainty (as can be seen from the central
plot of the bottom panel of Fig.~\ref{fig:transversity}) and a relatively
good improvement of $u$-quark (up to 80\% reduction of errors) as
well. It happens because of a higher statistics on the effective
neutron target in comparison to the proton target. The right plot of the bottom panel of
Fig.~\ref{fig:transversity} shows that in the kinematical region of 
SoLID, $0.05<x<0.6$, the errors will be reduced by approximately 90\%, i.e. one order of magnitude, for both
$u$- and $d$-quark transversities {\it if} measurements are
performed on both the proton and effective neutron targets.

Notice that the maximal improvements are attained in region covered by
the SoLID data $0.05<x<0.6$ and the impact decreases outside of this
region as expected.
One may notice the ``bump" around $x\simeq 0.2$ of the $d$-quark
transversity in all three bottom plots.  It appears to be an artifact
of usage of Soffer positivity bound~\cite{Soffer:1994ww} in the parametrization of transversity
for $u$- and $d$-quarks. Indeed, around $x\simeq 0.2$ the error corridor
saturates the bound and it shows up as a ``bump" in the ratio plot.

\begin{table}
\caption
{ Table of tensor charges computed using Eq.~(\ref{eq:trunctensor}).
Tensor charges are calculated at  $Q^2 = 2.4$ GeV$^2$ and $Q^2 = 10$
GeV$^2$ and in four regions of $x$ corresponding to the acceptance of
SoLID, $0.05 < x < 0.6 $; the full region,  $0 < x < 1$; and the regions
outside of acceptance, $0    < x < 0.05$, $0.6 < x < 1$. The errors
are computed at 90\% C.L. The isovector nulceon tensor charge  $g_T$ is calculated using the full region  $0 < x < 1$ and a truncated region $0.05 < x < 0.6 $, see Eq.~\eqref{eq:tensor}.  \label{table:tensor} }
\centering
\begin{tabular}{lllllc} 
observable
& $Q^2({\rm GeV^2})$
& ${\rm KPSY15}$
& $\delta_{\rm KPSY15}$ 
& $\delta_{\rm SoLID}$  
& $\delta_{\rm SoLID}/\delta_{\rm KPSY15}(\%)$  
\\\hline\hline
$\delta u^{[0    , 0.05 ]}$ & 2.4 & 0.046    & 0.010   & 0.005   & 49\\
$\delta u^{[0.05 , 0.6  ]}$ & 2.4 & 0.349    & 0.122   & 0.015   & 12\\
$\delta u^{[0.6  , 1    ]}$ & 2.4 & 0.018    & 0.007   & 0.001   & 14\\
$\delta u^{[0    , 1    ]}$ & 2.4 & 0.413    & 0.133   & 0.018   & 14\\
\hline
$\delta u^{[0    , 0.05 ]}$ & 10  & 0.051    & 0.011   & 0.005   & 46\\
$\delta u^{[0.05 , 0.6  ]}$ & 10  & 0.332    & 0.117   & 0.014   & 12\\
$\delta u^{[0.6  , 1    ]}$ & 10  & 0.0126   & 0.0048  & 0.0007  & 14\\
$\delta u^{[0    , 1    ]}$ & 10  & 0.395    & 0.128   & 0.018   & 14\\
\hline
$\delta d^{[0    , 0.05 ]}$ & 2.4 & -0.029   & 0.028   & 0.003   & 10\\
$\delta d^{[0.05 , 0.6  ]}$ & 2.4 & -0.200   & 0.073   & 0.006   & ~9\\
$\delta d^{[0.6  , 1    ]}$ & 2.4 & -0.00004 & 0.00009 & 0.00001 & 13\\
$\delta d^{[0    , 1    ]}$ & 2.4 & -0.229   & 0.094   & 0.008   & ~9\\
\hline
$\delta d^{[0    , 0.05 ]}$ & 10  & -0.035   & 0.030   & 0.003   & 10\\
$\delta d^{[0.05 , 0.6  ]}$ & 10  & -0.184   & 0.067   & 0.006   & ~9\\
$\delta d^{[0.6  , 1    ]}$ & 10  & -0.00002 & 0.00006 & 0.00001 & 14\\
$\delta d^{[0    , 1    ]}$ & 10  & -0.219   & 0.090   & 0.008   & ~9\\
\hline
 $g_{T}^{(\rm truncated)}$  & 2.4 & 0.55   & 0.14  & 0.018   & 13\\
 $g_{T}^{(\rm full)}$ & 2.4  & 0.64   & 0.15   & 0.021   & 14\\
\hline
 $g_{T}^{(\rm truncated)}$  & 10  & 0.51   & 0.13  & 0.017   & 13\\
 $g_{T}^{(\rm full)}$ & 10  & 0.61   & 0.14   & 0.020   & 14\\
\hline
\end{tabular}
\end{table}


The tensor charges can be calculated using Eq.~(\ref{eq:trunctensor})
if one neglects sea-quark contributions. In Table~\ref{table:tensor}
we present the estimated  improvements for the truncated tensor
charges at $Q^2 = 2.4$ GeV$^2$ and $Q^2 = 10$ GeV$^2$ separated into
three kinematical regions of $x$: the region of SoLID acceptance
($0.05<x<0.6 $) and the regions outside of SoLID coverage.  For the
region where SoLID has the maximum impact  we find the improvement of
about 90\% (up to one order of magnitude) for both $u$- and $d$-quark tensor charges.

\begin{figure}[htbp]
\centering 
\includegraphics[width=0.6\textwidth]{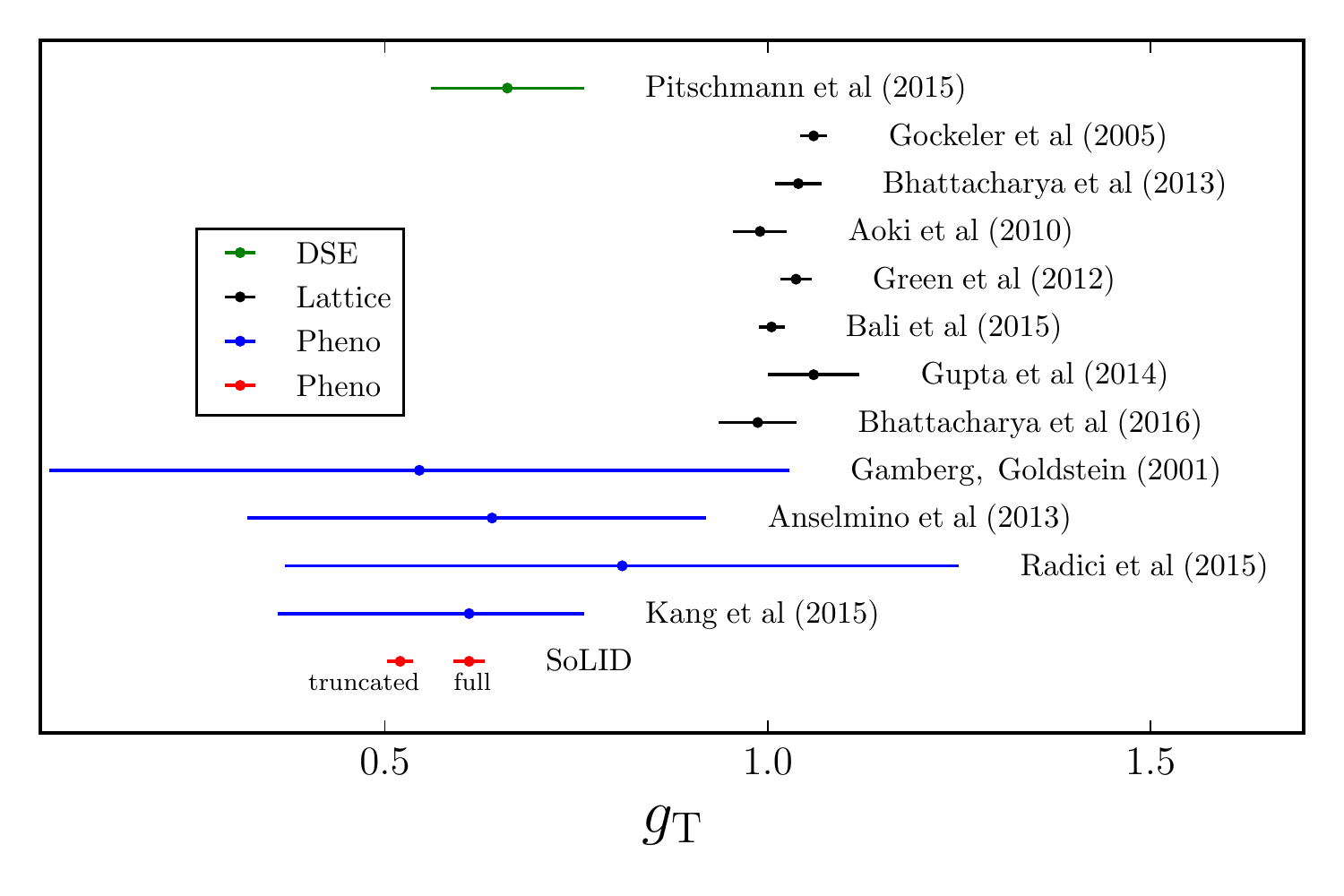}  
\vskip -0.3cm 
\caption{
The isovector nucleon tensor charge $g_T$  after the
pseudo-data of SoLID is taken into account is compared with result of Kang et al
2015~\cite{Kang:2015msa} at $Q^2=10$ GeV$^2$,  result from
Ref.~\cite{Radici:2015mwa}  (Radici et al 2015) at 68\% C.L. and
$Q^2=4$ GeV$^2$, and result from Ref.~\cite{Anselmino:2013vqa} at 95\%
C.L. (Anselmino et al 2013) at $Q^2=0.8$ GeV$^2$, and Ref.~\cite{Gamberg:2001qc} (Gamberg, Goldstein 2001) at  $Q^2=1$ GeV$^2$. 
Other points are
lattice computation  at $Q^2=4$ GeV$^2$ of Bali et al
Ref.~\cite{Bali:2014nma},  Gupta et al Ref.~\cite{Gupta:2015tpa},
Green et al Ref.~\cite{Green:2012ej}, Aoki et al
Ref.~\cite{Aoki:2010xg}, Bhattacharya et al
ref.~\cite{Bhattacharya:2013ehc,Bhattacharya:2016zcn}, Gockeler et al
Ref.~\cite{Gockeler:2005cj}. Pitschmann et al is DSE calculation
Ref.~\cite{Pitschmann:2014jxa} at $Q^2=4$ GeV$^2$.  Two SoLID points are the truncated and full tensor charges from Eq.~\eqref{eq:tensor}.}
\label{fig:comparison_gt}
\end{figure}

  Finally we present our estimates for the precision of extraction of
isovector nucleon tensor charge $g_T= \delta u - \delta d$,  after the data of SoLID is taken into account:
 \begin{eqnarray}
 g_{T}^{(\rm truncated)} =  +0.55^{+0.018}_{-0.018} \ , 
 \quad \quad \quad
 g_{T}^{(\rm full)} =  +0.64^{+0.021}_{-0.021} \ , 
 \label{eq:tensor}
\end{eqnarray}
at  $Q^2=2.4$ GeV$^2$ where \emph{truncated} means contribution from the region covered by
the SoLID data $0.05<x<0.6$, and \emph{full} is the contribution from $0<x<1$. See Table~\ref{table:tensor} for a detailed comparison.
The {\it precision} of this result can be readily compared to
precision of the lattice QCD calculations.  
As studied in Ref.~\cite{Radici:2015mwa}, parametrizations of transversity that are substantially different in the region not covered by experimental data  but similar in the region covered by the data 
lead to the growth of uncertainties of $g_T$ in the full kinematical region $0<x<1$. The relative improvement of the error, however, is less sensitive to the particular choice of parametrization especially in the region where data exists. With this in mind our result of the improvement of $g_{T}^{(\rm truncated)}$ is a more reliable estimate. As one can see from Eq.~\eqref{eq:tensor} and Fig.~\ref{fig:comparison_gt} we predict an order of magnitude improvement of the error. Future data from Electron Ion Collider will extend the region of the data and allow to explore low-x region. 

In Fig.~\ref{fig:comparison_gt} we compare our result with extraction
of Radici et al Ref.~\cite{Radici:2015mwa} at  $Q^2=4$ GeV$^2$,
Anselmino et al  Ref.~\cite{Anselmino:2013vqa} at $Q^2=0.8$ GeV$^2$; Gamberg, Goldstein 2001 Ref.~\cite{Gamberg:2001qc}  at  $Q^2=1$ GeV$^2$.
Our result is also compared to a series of lattice computations, at $Q^2=4$ GeV$^2$ of Bali et al
Ref.~\cite{Bali:2014nma},  Gupta et al Ref.~\cite{Gupta:2015tpa},
Green et al Ref.~\cite{Green:2012ej}, Aoki et al
Ref.~\cite{Aoki:2010xg}, Bhattacharya et al
Refs.~\cite{Bhattacharya:2013ehc,Bhattacharya:2016zcn}, Gockeler et al
Ref.~\cite{Gockeler:2005cj}. Pitschmann et al \cite{Pitschmann:2014jxa} is a DSE calculation at
$Q^2=4$ GeV$^2$.
The value of $g_T$ extracted from the data may influence 
searches beyond the standard model
\cite{DelNobile:2013sia,Bhattacharya:2011qm,Courtoy:2015haa}.

\section{Summary and Conclusions}
We have studied impact of future SoLID data on both the proton and
the effective neutron targets on extraction of transversity for $u$-
and $d$-quarks and tensor charge of the nucleon. 
A new method based on Hessian error analysis was developed
in order to estimate the impact of future new data sets on TMD
distributions.
Based on the global QCD  analysis with TMD evolution of the current data  of
Ref.~\cite{Kang:2015msa} we estimated that the combination of both the
proton and the effective neutron targets is essential for the
appropriate extraction of tensor charge.  As one can clearly see in
Fig.~\ref{fig:transversity} we predict a balanced improvement in the precision of extraction for both
$u$- and $d$-quarks up to one order of magnitude in the range $0.05<x<0.6$ with such a combination of measurements. 

We would like to emphasize that it is also important to investigate
other possible contributions to asymmetries that may influence
extraction of the quark transversity distributions from the
experimental data. One particular example is the higher-twist
contributions, which can be thoroughly studied when the future data
are available from Jefferson Lab 12 GeV upgrade, including both
spin-averaged and spin-dependent cross section measurements. In
addition, with the wide kinematic coverage in $Q^2$, the planed
Electron Ion Collider will provide valuable information on higher
twist contributions as well.     

Under assumptions of Ref.~\cite{Kang:2015msa} we also predict an
impressive improvement in the extraction of tensor charge as can be
seen in Table~\ref{table:tensor} in the presence of SoLID
measurements.  It appears that the  acceptance region of SoLID will reveal
most of contribution from $u$ and $d$ quarks to the tensor charge of the nucleon.  The contribution from the
region of high-$x$ not covered by SoLID ($x>0.6$) appear to be small
for both $u$ and $d$ quarks, see Table~\ref{table:tensor}. The same
is true for the contribution from low-$x$ region, ($x<0.05$).
The contribution to the tensor charge from anti-quarks at low-$x$
region was omitted in the present analysis. We leave for future the
study of the impact of the Electron Ion  Collider on the sea-quark transversity 
distributions.

The precision at which isovector tensor charge $g_T$ can be extracted
from the SoLID data will be comparable to the precision of lattice QCD
calculates, as can be seen from Fig.~\ref{fig:comparison_gt}, and
will provide a  unique opportunity for searches beyond the standard model.
Our results demonstrate the powerful capabilities of future measurements of
SoLID apparatus at Jefferson Lab 12 GeV Upgrade.

\section*{Acknowledgments}
We are grateful to Leonard Gamberg and John Arrington for useful
discussions.  This work was partially supported by the U.S.\
Department of Energy under Contract No.~DE-AC05-06OR23177 (A.P., N.S.,
J.C.),~DE-AC02-06CH11357 (Z.Y.),~DE-FG02-94ER40818 (K.A.),
No.~DE-AC02-05CH11231 (F.Y.), No.~DE-AC52-06NA25396 (Z.K.),
DE-FG02-03ER41231 (H.G., Z.Y., T.L.), by the National Science Foundation
under Contract No. PHY-1623454 (A.P.), and by the National Natural Science
Foundation of China under Contract No.~11120101004 (H.G., Z.Y., T.L.).

\section*{References}

\bibliographystyle{elsarticle-num}
\bibliography{\BibPath/solid}
 
\end{document}